\documentclass[aps,prl,twocolumn,superscriptaddress,showpacs]{revtex4}

\usepackage{epsf}

\begin{document}

\title{Absence of Edge Localized Moments in the Doped Spin-Peierls System
CuGe$_{1-x}$Si$_{x}$O$_3$}

\author{J. Kikuchi}
\email[E-mail:~]{kikuchi@ph.noda.sut.ac.jp}
\affiliation{Department of Physics, Science University of Tokyo,
Yamazaki, Noda, Chiba 278-8510, Japan}
\author{T. Matsuoka}
\affiliation{Department of Physics, Science University of Tokyo,
Yamazaki, Noda, Chiba 278-8510, Japan}
\author{K. Motoya}
\affiliation{Department of Physics, Science University of Tokyo,
Yamazaki, Noda, Chiba 278-8510, Japan}
\author{T. Yamauchi}
\affiliation{Institute for Solid State Physics,
University of Tokyo, Kashiwanoha, Kashiwa, Chiba 277-8581, Japan}
\author{Y. Ueda}
\affiliation{Institute for Solid State Physics,
University of Tokyo, Kashiwanoha, Kashiwa, Chiba 277-8581, Japan}

\date{August 2 2001}

\begin{abstract}
We report the observation of nuclear quadrupole resonance (NQR)
of Cu from the sites near the doping center in the spin-Peierls system
CuGe$_{1-x}$Si$_{x}$O$_3$.  The signal appears as the satellites in
the Cu NQR spectrum, and has a suppressed nuclear spin-lattice relaxation
rate indicative of a singlet correlation rather than an enhanced magnetic
correlation near the doping center.  Signal loss of Cu nuclei with no
neighboring Si is also observed.  We conclude from these observations that the
doping-induced moments are not in the vicinity of the doping center
but rather away from it.
\end{abstract}


\maketitle

Impurity-induced magnetism in low-dimensional quantum spin systems has been
the subject of recent intense theoretical and experimental studies. 
The discovery of the first inorganic spin-Peierls (SP) material
CuGeO$_3$ \cite{hase93} has opened up the field of experimental research
into the impurity effects on the SP transition which is inherent in the
spin-$\frac{1}{2}$ Heisenberg antiferromagnetic (AF) spin chains
coupled with phonons.  In-chain substitution or doping of Zn, Mg, Ni
for Cu \cite{oseroff95,hase95,masuda98,grenier98} as well as off-chain
substitution of Si for Ge \cite{grenier98,renard95} have been studied
intensively, and the temperature-concentration magnetic phase diagrams
are now becoming established.  Independently of the types of doping, a
rapid decrease of the transition temperature $T_{\rm SP}$ followed by
a disappearance of SP long-range order (LRO) is commonly observed.  At
the same time, there appears a new AF phase at low temperatures
coexisting with SP-LRO at very small dopant concentrations.  The
coexistence of SP- and AF-LRO has attracted much attention because
they had been thought to be exclusive.

The AF state induced by doping is rather unconventional. Although the
AF order is commensurate, the size of an ordered
moment is small and spatially varying \cite{kojima97}.  It is widely
believed that the inhomogeneous AF-LRO results from the enhanced
staggered correlation or the appearance of staggered moments near the
doping center, of which existence is suggested from numerical
calculations on a dimerized chain with open boundaries
\cite{martins97,laukamp98,watanabe99}, phase-Hamiltonian approach
assuming reduced dimerization near the impurity \cite{fukuyama96}, and
a soliton binding to the impurity due to interchain elastic couplings
\cite{khomskii96}.  However, there is by now no experimental evidence
that the staggered moments are really induced near the doping center
in the SP system.

In this Letter, we present Cu NQR evidences that this is not the case
in CuGeO$_3$ doped with Si.  The enhanced moments appear not in the
vicinity of Si atoms segmenting a linear chain but {\it in between} them,
which comes from the flexibility of the lattice locating singlet bonds
on the edges of an open segment.

Single crystals of CuGe$_{1-x}$Si$_{x}$O$_3$ were grown by a floating
zone method.  The Cu NQR spectrum was taken under zero external field
by integrating the spin-echo signal while changing the frequency point
by point.  Nuclear spin-lattice relaxation rate was measured by an
inversion recovery method.

\begin{figure}
\epsfxsize=85mm 
\centerline{\epsfbox{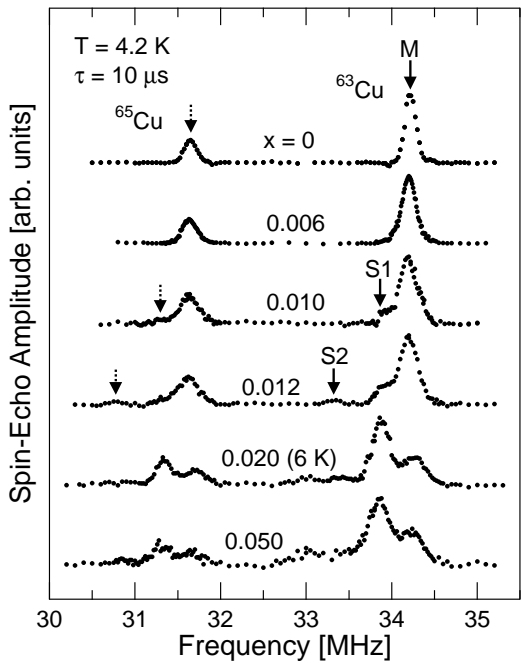}}
\caption{Cu NQR spectrum in CuGe$_{1-x}$Si$_{x}$O$_3$ at 4.2 K
($x=0.02$ at 6 K).  The intensities are normalized by the signal
maximum for each $x$.  M, S1 and S2 with arrows (solid: $^{63}$Cu,
dotted: $^{65}$Cu) denote the main line, satellite 1 and satellite 2,
respectively (see text).}
\label{spectrum}
\end{figure}

The Cu NQR spectra in the SP ($x$ = 0, 0.006, 0.010, 0.012) and uniform ($x$
= 0.02, 0.05) phases of CuGe$_{1-x}$Si$_{x}$O$_3$ are shown in Fig.\
\ref{spectrum}.  As the Si content $x$ increases, satellite lines
appear in addition to the original $^{63,65}$Cu NQR lines similarly
for both isotopes.  From the isotopic ratios of resonance frequencies and
intensities between the satellites, we associate the satellites with
Cu sites having different crystallographic environment from that in
the non-doped sample.  Based on the intensity argument below, the
satellites can be assigned as the signal from Cu nuclei with Si on the
first- and/or second-neighbor Ge sites.  Local change of lattice
dimerization due to Si substitution is unlikely for the appearance of
the satellites because the Cu NQR frequency in CuGeO$_3$ is known to
be insensitive to the lattice dimerization\cite{kikuchi94,itoh95}.

\begin{figure}
\epsfxsize=90mm
\centerline{\epsfbox{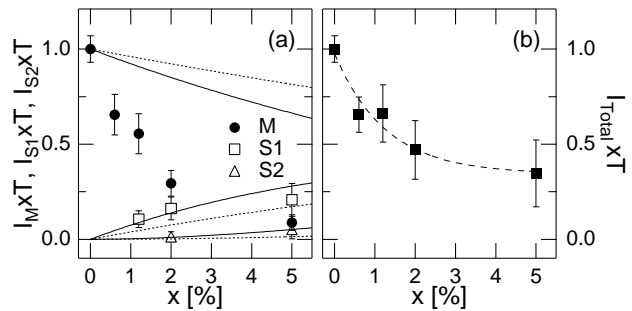}}
\caption{Si concentration dependence of the integrated intensity of the
$^{63}$Cu NQR spectrum in CuGe$_{1-x}$Si$_{x}$O$_3$ multiplied by
temperature.  The intensities are normalized by the number of
$^{63}$Cu nuclei in the $x=0$ sample.  The intensity of each NQR line
(a), and the total intensity (b).  The dotted and solid lines in (a)
are $_{n}B_{k}$ ($k=0$, 1, 2) with $n=4$ and 8, respectively.  The
dashed line in (b) is a guide to the eyes.}
\label{intensity}
\end{figure}

We restrict the following discussion to the $^{63}$Cu NQR. For convenience, we
call the NQR line observed in the non-doped sample as the main (M)
line.  The satellite in the vicinity of the line M is called satellite
1 (S1) and the weaker one at lower frequencies the satellite 2 (S2). 
To compare the number of nuclei which contributes to each NQR line, we
deconvoluted the spectrum by fitting to the sum of Gaussians. 
Figure\ \ref{intensity}(a) shows the $x$ dependence of the integrated
intensities $I_{\rm M}$, $I_{\rm S1}$ and $I_{\rm S2}$ of the lines M,
S1 and S2 multiplied by temperature $T$ which are proportional to the
number of nuclei contributing to the NQR line\cite{abragam61}.  The integrated
intensities were extrapolated to $\tau$ = 0 ($\tau$ being the
radio-frequency pulse separation) by measuring the spin-echo decay at
the peak position of the lines M, S1 and S2, and were normalized by the number
of $^{63}$Cu nuclei included in each sample.  We find a rapid decrease of
$I_{\rm M}\cdot T$ and a moderate increase of $I_{\rm S1}\cdot T$ and
$I_{\rm S2}\cdot T$ with $x$.  Shown in Fig.\ \ref{intensity}(b) is
the integrated intensity of all the NQR lines $I_{\rm total}=I_{\rm
M}+I_{\rm S1}+I_{\rm S2}$ times $T$ as a function of $x$. 
Surprisingly enough, $I_{\rm total}\cdot T$ decreases with increasing
$x$ in spite of the fact that we do not dilute the Cu site.  This
unambiguously demonstrates that a part of Cu nuclei is lost from NQR
observation.

There seem to be two possible origins for the loss of NQR signal
in CuGe$_{1-x}$Si$_{x}$O$_3$.  (i) Inhomogeneity of the
electric-field gradient (EFG): Since an NQR frequency is highly sensitive to
the local charge distribution, the signal may be wiped out if Si
substitution introduces significant inhomogeneity of the EFG at the Cu
site leading to a spread of NQR frequencies.  It is expected that this
is most effective, if any, near the doping center.  (ii) Appearance of
strongly-enhanced moments: Si substitution for Ge magnetically
activates some of Cu$^{2+}$ spins which otherwise is in a
spin-singlet state.  The signal cannot be observed if the nuclear
spins on such Cu sites relax in a time shorter than the time domain of a
pulsed NQR experiment ($>$ a few $\mu$s).  It is difficult to
predict the location of the induced moments in general, because it
depends on the microscopic mechanisms for their appearance.

In order to know which part of the Cu nuclei becomes unobservable, we
model the Si concentration dependence of the integrated intensity
based on the probability argument.  We assume that the NQR
frequency is determined by the number of Si atoms on the first- and/or
second-neighbor Ge sites for Cu.  The others are neglected because
they are relatively far apart.  The satellite 1(2) hence corresponds
to the Cu sites with one (two) Si atom(s) among four or eight
neighboring Ge sites, and the main line to the Cu site without Si
neighbors which is equivalent to the one in the non-doped sample.  The
probability of finding Cu atoms with $k$ (= 0, 1, 2) Si among $n$ (= 4
or 8) neighboring Ge sites is nothing but the binomial probability
$_{n}B_{k}(x)$ = $_{n}C_{k}\,x^{k} (1-x)^{n-k}$.

In Fig.\ \ref{intensity}(a) we compared $_{4}B_{k}(x)$ and $_{8}B_{k}(x)$
($k=0$, 1, 2) with $I_{\rm M}\cdot T$, $I_{\rm S1}\cdot T$ and $I_{\rm
S2}\cdot T$.  It is apparent that the satellite intensities are best
reproduced by taking $n=8$.  On the other hand, $I_{\rm
M}\cdot T$ decreases much faster than $_{8}B_{0}(x)$.  If we try to
fit $I_{\rm M}\cdot T$, we must take $\sim$40 Ge atoms up to the
tenth-neighbor Ge sites.  This, of course, cannot reproduce $I_{\rm
S1}\cdot T$ and $I_{\rm S2}\cdot T$ and is quite unreasonable for the
following reason.  If the Si substitution effect so long-ranged, the
satellite is considered to be severely broadened and may be wiped out
because there are so many nonequivalent Si configurations
corresponding to that satellite\cite{fn:config}.  In fact, we can
observe the satellites with relatively narrow linewidths which
contradicts with the above expectation.  We therefore exclude the
possibility that the rapid decrease of $I_{\rm M}\cdot T$ is due to
long-range nature of the Si substitution effect on the EFG.

It is now clear that the signal loss occurs for the Cu sites {\it without}
neighboring Si.  This readily exclude the EFG inhomogeneity scenario
for the loss of signal because there seems no disturbance to the EFG at
such Cu sites.  More importantly, local moments which appear in the
vicinity of the impurity are clearly incompatible with the fact that
the Cu sites near Si are observable as the satellites.  A consistent
explanation for the loss of signal from the main line would be the
appearance of enhanced moments away from the doping center. 
Considering the effective interruption of linear chains by
off-chain substitution \cite{khomskii96}, this implies, contrary to a
common belief, that the local moments are induced in the middle of a
segment rather than near the open edges.
\begin{figure}
\epsfxsize=82mm 
\centerline{\epsfbox{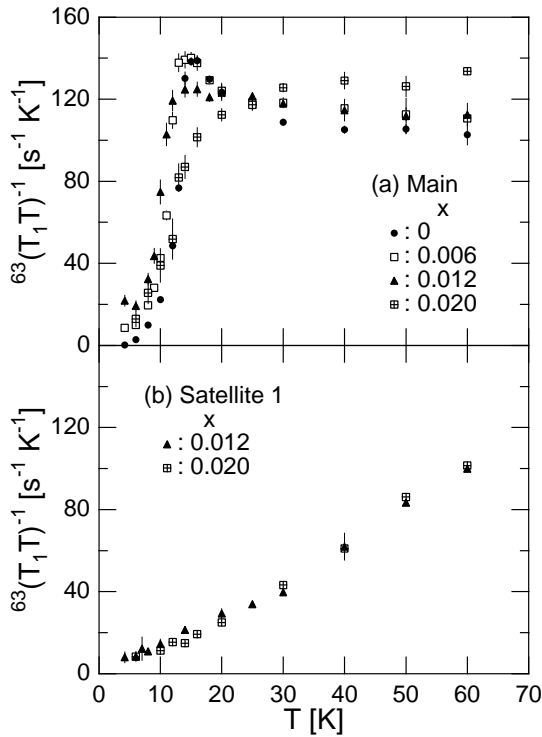}}
\caption{Temperature dependence of the $^{63}$Cu nuclear spin-lattice
relaxation rates $^{63}1/T_ 1 T$ for the line M (a), and the line S1 (b) in
CuGe$_{1-x}$Si$_{x}$O$_3$.}
\label{T1TvsT}
\end{figure}

The absence of impurity-induced moments near the doping center is also
supported by the nuclear spin-lattice relaxation rate of Cu. 
Non-exponential recovery of Cu nuclear magnetization $M(t)$ 
was observed at temperatures as high as 40 K for $x > 0$ indicating
spatially-nonuniform relaxation processes.  We analyzed the recovery
curves by fitting to the stretched exponential form $p(t) \equiv
1-M(t)/M(\infty) \propto \exp[-t/T_1-(t/\tau_1)^{1/2}]$ which works
well in the system with dilute magnetic impurities\cite{mchenry72}. 
In the present case, $1/T_1$ can be interpreted as the relaxation due to
the on-site electronic spin via the hyperfine coupling and $1/\tau_1$
the relaxation resulting from the impurity-induced moments via the
much weaker classical dipolar coupling.  Figure\ \ref{T1TvsT} shows
the $T$ dependence of $^{63}$Cu nuclear spin-lattice relaxation rates
$^{63}1/T_ 1 T(\rm M)$ and $^{63}1/T_ 1 T(\rm S1)$ for the lines M and
S1\cite{fn:T1NQR}.  It is evident from comparison of the absolute
values of $^{63}1/T_ 1 T$ that Cu$^{2+}$ spins near Si are
magnetically less active than those without Si nearby.

We can trace the size effect on the dynamics of an open SP chain
via $1/T_ 1 (\rm M)$ and $1/T_ 1(\rm S1)$ which probe bulk- and edge-spin
fluctuations, respectively.  Let us look at $1/T_ 1 (\rm M)$ first. 
In the uniform (U) phase of CuGeO$_3$, the $T$ dependence of $1/T_ 1 T$ at the
Cu site is understood as being contributed by the nearly
$T$-independent uniform (wave vector $q=0$) mode and the staggered
($q=\pi$) mode showing $T^{-1}$-like divergence at low $T$\cite{itoh95}.  As
shown in Fig.\ \ref{T1TvsT}(a), the low-$T$ increase of $^{63}1/T_ 1 T
(\rm M)$ due to the staggered contribution is rounded off by doping
and is completely suppressed in the $x=0.02$ sample in spite of the
fact that the wave vector $q=\pi$ characterizes the low-$T$ AF-LRO.
In addition, $^{63}1/T_ 1 (\rm M)$ does not exhibit critical divergent
behavior typical for three-dimensional magnetic ordering even at low
$T$.  These observations suggest that the observed bulk spins do not
play an active role for the appearance of AF-LRO. In the SP state ($x
\leq 0.012$), $^{63}1/T_ 1 T(\rm M)$ is increased by doping.  This may
result from the reduction of the SP gap and/or the appearance of
spin-wave-like excitations below the SP gap\cite{hirota98}.  Such an
$x$ dependence of the relaxation rate is qualitatively the same as
that observed in Mg-doped CuGeO$_3$\cite{yitoh01}.

As a clear contrast, $^{63}1/T_ 1 T (\rm S1)$ is essentially $x$-independent
which indicates that the dynamics of the edge spin is insensitive to not only
the size of a segment but also the existence of SP-LRO. The most striking
feature of the result is the large suppression of $^{63}1/T_ 1 T (\rm
S1)$ from the bulk $^{63}1/T_ 1 T(\rm M)$ which begins gradually below
about 60 K.  Pouget and his co-workers have revealed from the x-ray diffuse
scattering studies that in-chain pretransitional lattice fluctuations
or local dimer correlations exist as high as 40 K in both pure and
Si-doped CuGeO$_3$ \cite{pouget94,pouget01}.  The suppression of the
edge-spin fluctuations probed by $^{63}1/T_ 1 T (\rm S1)$ may be
associated with such preformation of singlet dimers on the edges.
\begin{figure}
\epsfxsize=75mm 
\centerline{\epsfbox{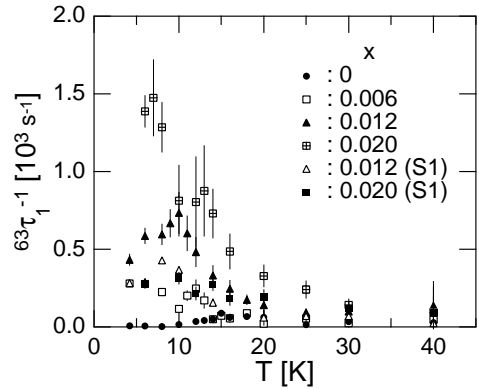}}
\caption{Temperature dependence of the relaxation rate $^{63}1/\tau_1$
due to impurity-induced moments in CuGe$_{1-x}$Si$_{x}$O$_3$.}
\label{tau1vsT}
\end{figure}

Figure\ \ref{tau1vsT} shows the $T$ dependence of the relaxation rate
$^{63}1/\tau_1$ due to the impurity-induced moments.  Minor differences of
$^{63}1/\tau_1$ between the lines M and S1 are probably due to
long-range nature of the dipolar coupling.  $^{63}1/\tau_1$ is
enhanced by doping and exhibits a Curie-Weiss-like increase at low $T$
but above $T_{\rm SP}$.  The low-$T$ increase of $^{63}1/\tau_1$
indicates slowing down of the impurity-induced moments, because
$1/\tau_1$ is roughly proportional to the correlation time of the
impurity-induced moments\cite{mchenry72}.

Now that the absence of doping-induced moments next to an impurity or an
open chain end becomes obvious, theories predicting local moments near the
chain end are not relevant to describe the magnetic properties of
Si-doped CuGeO$_3$.  Though gradual, there is a clear tendency of the
edge spins to become nonmagnetic much faster than the bulk spins when
lowering temperature in the U phase.  This suggests that in the SP
phase, strong bonds in which spins experience most of the time in a
singlet state are formed on the open edges.  The key ingredient will
be a coupling with phonons which can adjust dynamically a dimerization
pattern so as to minimize the total energy of the spin and the lattice
systems.  It has recently been shown numerically that the lattice
relaxation stabilizes the dimerization pattern with the strong bonds
on the edges of an open chain \cite{hansen99,onishi00}, although such
a possibility has been pointed out from intuitive considerations
\cite{laukamp98,watanabe99,khomskii96}.  The dimerization in Si-doped
CuGeO$_3$ should therefore be such that an open chain is
``terminated'' by the strong bonds.  A defect of bond alternation or a
soliton will then be formed in the middle of a segment with an odd
number of sites, accompanying modulation of lattice dimerization and
staggered polarization of magnetic moments around it.  This latter
causes a dramatic loss of NQR signal from the Cu sites with no
neighboring Si depending on the soliton size in comparison with the
size of a segment.  Similar discussion may be applied to the system
with in-chain nonmagnetic dopants such as Zn and Mg because the
interruption of a chain is more direct.  It is worth noting that the
signal loss may occur in the even-numbered segment because of the
formation of a soliton-antisoliton pair at finite temperatures
\cite{hansen99,laukamp98}.  This may be an origin for the
more-than-half reduction of the intensity of the line M at higher
doping levels from the model calculation.

The present $^{63}$Cu relaxation data show that the local singlet correlation
persists in the sample with no SP-LRO. This renders support for the
soliton picture of the SP transition in pure and doped CuGeO$_3$ that
the SP-LRO disappears when the interchain coherence of the lattice distortion
is lost rather than the dimerization itself \cite{khomskii96,pouget01}. 
The rapid decrease of $T_{\rm SP}$ due to doping may result from the
weakness of the interchain elastic coupling leaving solitons away from
the impurities in spite of the ``irregular'' dimerization pattern
of that segment \cite{khomskii96}.  As for AF-LRO, nonmagnetic nature of
the edge site, suppressed bulk AF correlations and the absence of
critical divergence of $1/T_ 1$ suggest that both the edge and bulk
spins play a passive role for magnetic ordering.  What becomes
critical is the fluctuation of impurity-induced moments in the middle
of a segment, which allows one to view the AF-LRO in Si-doped
CuGeO$_3$ as arising from the interchain magnetic coupling between the
impurity-induced moments.

We thank for valuable discussion with Y. Itoh and C. Yasuda.  T. Fujii and S.
Ishiguro are also acknowledged for experimental help.


\begin{references}
\bibitem{fn:config} There are 10 and 55 nonequivalent configurations of
Si for the lines S1 and S2, respectively, if we take account of up to
the tenth-neighbor Ge sites, whereas we have only 2 (S1) and 4 (S2)
configurations in the present model.
\bibitem{fn:T1NQR} $T_1$ is defined as the time constant
characterizing the exponentially-decaying part of $p(t)$
measured by NQR as described in the text.  In Refs.\cite{kikuchi94} and
\cite{itoh95}, on the other hand, $T_1$ is defined as $p(t) \propto
\exp(-3t/T_1)$ in which the factor 3 is incorrect for the nonaxial EFG
in CuGeO$_3$.
\end{references}

\end{document}